\documentstyle[floats,aps,epsf]{revtex}
\begin{document}
\title{Cluster structure in stable and unstable nuclei}
\author{Y. Kanada-En'yo}
\address{Institute of Particle and Nuclear Studies,\\
High Energy Accelerator Research Organization, \\
Oho 1-1, Tsukuba-shi 305-0801, Japan}
\author{M. Kimura}\address{Institute of Physical and Chemical Research(RIKEN),
 Saitama 351-0198, Japan}
\author{H. Horiuchi}\address{Department of Physics, Kyoto University,\\
Kitashirakawa-Oiwake, Sakyo-ku, 
Kyoto 606-01,
Japan}
\maketitle
%
%
\abstract{Cluster structure in stable and unstable nuclei has been studied.
We report recent developments of theoretical studies on cluster aspect,
which is essential for structure study of light unstable nuclei.  
We discuss negative parity bands in even-even Be and Ne isotopes and
show the importance of cluster aspect. Three-body cluster structure 
and cluster crystallization are also introduced. It was found that 
the coexistence of cluster and mean-field aspect brings a variety of 
structures to unstable nuclei.}
%
} 
\section{Introduction}
\label{intro}
Clustering is one of the essential features in 
nuclear dynamics. As already known, cluster structures appear
in light stable nuclei such as $^8$Be, $^{12}$C and $^{20}$Ne.
Owing to recent developments of experimental and theoretical
studies on unstable nuclei, 
cluster structures have been found also in light unstable nuclei.
For instance, 
cluster states have been 
suggested in neutron-rich Be isotopes
\cite{TANIHATA,FREER,SAITO,SEYA,OERTZEN,ARAI,DOTE,ENYOg,ITAGAKI,OGAWA,ENYObe14,ENYObe12,ITO,DESCOUVEMONT}.
Furthermore, in the heavier nuclei, 
the importance of cluster aspect are found in 
such phenomena as molecular resonances, which has been observed 
in stable $sd$-shell and $pf$-shell nuclei.
On the other hand, we should remind the reader that 
the mean-field nature is the other essential aspect.
It is important that the coexistence of these two natures, 
the cluster and the mean-field aspects brings a variety of structure to
unstable nuclei as well as stable nuclei.

We can see the coexistence of cluster and mean-field aspects in such stable 
nuclei as $^{12}$C, where the 
3$\alpha$ cluster structure plays an important role.
The ground state is considered to contain the 3$\alpha$-cluster 
{\it and} $p_{3/2}$ sub-shell closure configurations\cite{Kan98}.
In the excited states above the 3$\alpha$ threshold energy, 
developed 3$\alpha$ cluster structures are expected. The
 $0^+_2$, $0^+_3$ and $2^+_1$ states of $^{12}$C have been discussed 
in relation to the 3$\alpha$ structure for a long time
\cite{Morinaga,Fujiwara80,Kan98,Tohsaki01,Funaki03}.

In unstable nuclei, a variety of cluster structure appears, and
the coexistence of cluster and mean-field aspects becomes further important.
In halo nuclei, $^6$He and $^{11}$Li, the behavior of 
valence neutrons is described by a hybrid configuration of 
the independent single-particle motion {\it and} 
di-neutron structure 
\cite{Varga94,Csoto93,Aoyama95,Tosaka90,Descouvemont97,Aoyama01}.
The former is a kind of mean-field nature, 
and the latter is regarded as cluster aspect.
In the neutron-rich Be isotopes, a molecular-orbital picture well works
to describe the structure of low-lying states
\cite{SEYA,OERTZEN,ITAGAKI,Oertzen97}.
2$\alpha$ core and valence neutron structure are found in   
many low-lying states of neutron-rich Be.
The molecular orbitals are formed 
in the mean-field of 2$\alpha$-cluster system, and 
the valence neutrons are moving in the molecular orbitals 
around the 2$\alpha$ core.  It means that
a kind of mean-field nature is seen in the valence neutron behavior,
and simultaneously the 2$\alpha$-cluster core plays an important role
to form the mean-field for the molecular orbitals. 
In the highly excited states of $^{12}$Be,  
molecular resonant states with $^6$He clusters have been suggested
\cite{ENYObe12,ITO,DESCOUVEMONT}. 

As mentioned above, a variety of structure has been revealed and it
motivates one to extend theoretical frameworks.
In these years, the development of theoretical framework for 
cluster structure has been 
remarkable following the progress of physics of unstable nuclei.
Such cluster models as core+neutrons models and multi-cluster models
have been applied to unstable nuclei.
These models are useful to describe the details of the 
relative motion between clusters and motion between core and valence neutrons.
In addition to simple cluster models based on two-body or three-body 
calculations, extended models such as stochastic variational method(SVM)
\cite{ARAI,OGAWA,Varga94,Varga95},
molecular orbital method(MO)\cite{SEYA,OERTZEN,ITAGAKI,OKABE} 
and generator coordinate method(GCM)
\cite{Descouvemont97,Descouvemont94,Descouvemont01,Descouvemont02}
have been developed for structure study of 
unstable nuclei.
We should give a comment on another important development of cluster models 
concerning studies of resonances in loosely bound systems. 
The complex scaling 
method(CSM)\cite{Csoto93,Aoyama95,Aoyama01,Kruppa,Aoyama97},
method of analytic continuation in the 
coupling constant(ACCC)\cite{Aoyama03}, and R-matrix theory\cite{Arai03} were
applied to estimate widths of the resonances in cluster models.

In most of cluster models, the existence of clusters is {\it a priori}
assumed. However, the assumption of clusters 
is not necessarily valid for systematic 
study of unstable nuclei. Instead, it is important to take into account
degrees of all single nucleons. In this sense, a method of antisymmetrized 
molecular dynamics(AMD) \cite{Kan98,ENYObc,ENYOsup,AMDrev} is one of
the powerful approaches which do not rely on the model assumption of cluster 
cores.
The wave function of the AMD is similar to the
Bloch-Brink model, but is based completely on single nucleons, and therefore
the degrees of all single-nucleon wave functions are independently 
treated.
Due to the flexibility of the AMD wave function, it can describe
various cluster states as well as shell-model-like states.
Similar model space has been adopted in 
a method of Fermionic molecular dynamics(FMD)\cite{Feldmeier95}. 
One of the remarkable 
advantages of the recent version of FMD is that 
the effect of tensor force is incorporated 
based on realistic nuclear forces in this framework \cite{Neff03}, while 
phenomenological effective nuclear forces are usually used in other cluster 
models.

Owing to the progress of calculations with these models, 
the structure study of stable and unstable nuclei
has been now extended to a wide mass number region up to
$sd$ and $pf$-shell region, and it reveals
the importance of cluster aspect in ground and excited states of various
nuclei. In this paper, we take some topics on cluster aspect in unstable
nuclei. In the next section \ref{sec:1}, 
we focus on the negative parity bands in 
even-even nuclei. The excited states of Be and Ne isotopes are discussed
in relation to cluster structure.
In section \ref{sec:2}, we report 3$\alpha$ structure in C isotopes
and discuss the mechanism of rigid cluster structure.
Finally, we give a symmary in \ref{sec:3}.

\section{Negative parity bands in even-even nuclei}
\label{sec:1}

As well known, $^{20}$Ne has a $^{16}$O+$\alpha$ cluster structure,
which was confirmed by parity doublets, $K^\pi=0^+_1$ and $K^\pi=0^-_1$
rotational bands. The parity doublets arise from the 
reflection asymmetry of the intrinsic state, which is caused by 
$^{16}$O+$\alpha$ clustering.
Thus, negative parity bands can be good probes
for cluster structure.
It is interesting that another type of negative parity rotational band
with cluster structure appears in neutron-rich nuclei 
based on single-particle excitation of neutron orbitals. 
In such a state, the origin of the negative parity is the 
single-particle excitation. 
A typical example is the $1p$-$1h$ excitation of 
the valence neutron orbitals in the molecular orbital states.
The negative parity bands with the $1p$-$1h$ excitation have been suggested in
low-lying states of neutron-rich Be isotopes.
One of the characteristics of those negative-parity
bands is the quanta $K^\pi=1^-$, which differs
from the $K^\pi=0^-$ of the parity doublets.
In this section, we discuss the negative parity bands in neutron-rich 
Be and Ne isotopes.

\subsection{Be isotopes}
\label{subsec:1}

The low-lying states of neutron-rich Be isotopes are well 
described by the molecular orbital picture based on the 2$\alpha$ core 
and valence neutron structure. In the 2$\alpha$ system, the molecular
orbitals are formed by a linear combination of $p$-orbits around the
2$\alpha$ core. In neutron-rich Be, the valence neutrons 
occupy the molecular orbitals.
The negative parity orbital are called as 
'$\pi$-orbitals' and the longitudinal orbital with positive parity is a
'$\sigma$ orbital'. As a result of the formation of molecular orbitals 
in Be isotopes, a negative parity $K^\pi=1^-$ band is constructed 
because of the one valence-neutron excitation of the 
molecular orbitals in the rotating cluster structure.
 In $^{10}$Be, the $K^\pi=1^-$ band is regarded as such a molecular band
with the one-neutron excitation, which can be described by
a $\pi^1\sigma^1$ configuration in the molecular orbital picture
\cite{OERTZEN,ENYOg,ITAGAKI,Oertzen97}.

\begin{figure*}
\epsfxsize=8cm
    \centerline{\epsffile{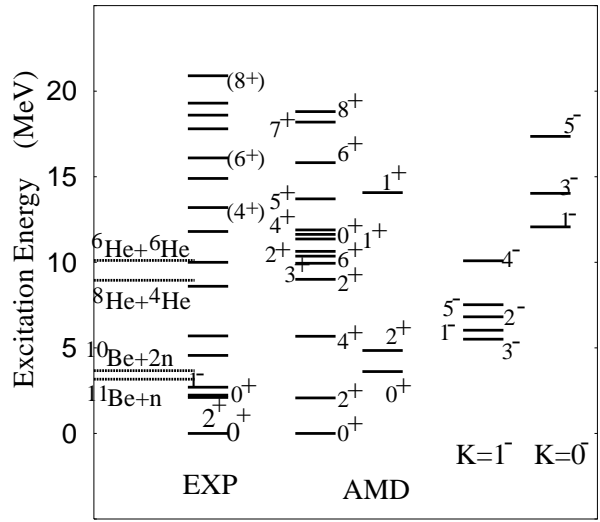}}
\caption{Energy levels of $^{12}$Be. The details of the AMD calculations
are explained in \protect\cite{ENYObe12}.}
\label{fig:1}       
\end{figure*}

In $^{12}$Be, many rotational bands are theoretically suggested.
Figure \ref{fig:1} shows the energy levels of $^{12}$Be calculated by 
variation after spin-parity projection in the AMD framework with
MV1 force\cite{TOHSAKI}. The details of the framework and calculations
are given in \cite{ENYObe12} and references therein.
The theoretical results obtained by the AMD calculations well agree to the
experimental data. We found three positive parity rotational bands 
$K^\pi=0^+_1, 0^+_2$ and $0^+_3$.
The ground band consists of the intruder states ($2\hbar\omega$
excited configurations), which are
well-deformed states with the 2$\alpha$ core in the surrounding neutrons.
On the other hand, the normal neutron-shell-closed states belong to the
$K^\pi=0^+_2$ band. It means that the breaking of neutron magic number $N=8$
occurs in $^{12}$Be. In the $K^\pi=0^+_3$ band, 
$^6$He+$^6$He molecule-like states are predicted in the results.
The experimentally measured $4^+$ and $6^+$ states are the candidates
of these molecular resonant states.

Next, let us analyze the negative parity states of $^{12}$Be.
In the negative parity states, two bands $K^\pi=1^-$ and $K^\pi=0^-$ 
are obtained by the AMD calculations. The lower one is the $K^\pi=1^-$
band which consists of $1^-$, $2^-$, $3^-$, $4^-$ and $5^-$ states.
These states are the molecular orbital states, which can be described by 
the $\pi^3\sigma^1$(three neutrons in the $\pi$-orbitals and
one neutron in the $\sigma$-orbital) configuration of the valence neutrons
in the 2$\alpha$ cluster system.
This $K^\pi=1^-$ band is consistent with the molecular orbital band predicted 
by von Oertzen et al.\cite{Oertzen97}. 
On the other hand, we also obtain the higher negative parity band($K^\pi=0^-$).
The $K^\pi=0^-$ band is formed by a parity asymmetric neutron structure
of the intrinsic state. This band is associated with the well-known 
parity doublet $K^\pi=0^-$ band in $^{20}$Ne.
In the GCM calculations of $^{12}$Be performed by Descouvemont and Baye
\cite{Descouvemont01},
a consistent result of the $K^\pi=0^-$ band was obtained by the
$^8$He+$\alpha$-like cluster structure.
Thus, the AMD results suggest the existence of two negative parity bands
$K^\pi=1^-$ and $0^-$. The $K^\pi=1^-$ band is the molecular orbital band,
while the $K^\pi=0^-$ band is the parity doublet caused by the 
parity asymmetric intrinsic state. Both bands have developed 
cluster structure,however, there exists a remarkable difference 
between these two bands concerning the origin of negative parity.
In the $K^\pi=1^-$ band, the negative parity originates from the one-particle
excitation of the valence neutron in the molecular orbitals. 
On the other hand, the negative parity of the $K^\pi=0^-$ band arises from 
the parity asymmetric shape of the intrinsic state. 
Due to the collectivity, the $E1$ transition into
the ground state is stronger as $B(E1;0^+_1\rightarrow 1^-_2)$=1.0 e$^2$fm$^4$
from the $K^\pi=0^-$ band than $B(E1;0^+_1\rightarrow 1^-_1)$=0.2 e$^2$fm$^4$
from the $K^\pi=1^-$ band.
It is interesting that the lower one is the 
molecular orbital $K^\pi=1^-$ band which has a kind of mean-field nature in
the valence neutron behavior.
We consider that the experimentally observed $K^\pi=1^-$ state\cite{IWASAKI} 
corresponds to
the band-head state the $K^\pi=1^-$ band.

\subsection{Ne isotopes}
\label{subsec:2}

In analogy of neutron-rich Be isotopes, von Oertzen proposed molecular orbital
structure of Ne isotopes based on the $^{16}$O+$\alpha$-cluster core
\cite{Oertzen97}.
In the molecular orbital picture, the cluster structure may develop
when the valence neutrons occupy the molecular 
$\sigma$-orbitals, which correspond to longitudinal $fp$-like
orbits.
In $^{22}$Ne, they proposed a developed cluster structure 
where two valence neutrons occupy the $\sigma$ orbit.
As a result of the development of cluster,
the parity doublet $K^\pi=0^-$ band arise from the parity
asymmetric structure of the $^{16}$O+$\alpha$-cluster core.
In the experiments, Rogachev et al. recently observed the $\alpha$-cluster 
states in negative parity bands which start from 
the $1^-$ state at 12 MeV excitation energy of 
$^{22}$Ne\cite{Rogachev}. These negative parity states can be described by the developed
cluster structure in the molecular orbital picture. 
The microscopic calculations of the excited states of $^{22}$Ne were
performed by Dufour and Descouvemont by using a GCM method within 
a $^{18}$O+$\alpha$ model\cite{Dufour}. 
The negative parity bands obtained by the GCM calculations well 
agree to the observed $\alpha$-cluster states.
These negative parity states in the parity doublet band
exist in the energy region above $^{18}$O+$\alpha$ threshold 
energy(9.7 MeV excitation energy of $^{22}$Ne).
One can expect that other negative parity bands with 
single-particle excitations of neutron orbits may appear
in a lower energy region.

Next, we discuss a further neutron-rich nucleus, $^{30}$Ne.
The AMD+GCM calculations of $^{30}$Ne 
have been performed, and many low-lying bands have been
predicted by Kimura et al.\cite{Kimura-ne}.
In the AMD calculations, they found that the
ground band of $^{30}$Ne is dominated by $2\hbar\omega$ configuration,
which indicates the breaking of magic number $N=20$.
The results suggest $4\hbar\omega$ state with a $4p$-$4h$ neutron 
configuration appear in the low-lying 
$0^+_3$ band(the excitation energy $E_x(0^+_3)=4$ MeV). 
The $4p$-$4h$ state has a 
parity asymmetric proton structure which indicates the developed 
$^{16}$O+$\alpha$-cluster core.
One of the striking results is the prediction of a low-lying 
$K^\pi=1^-$ band with a $3p$-$3h$ neutron configuration.
The structure of the $K^\pi=1^-$ band is described by 
the excitation of neutrons in the deformed mean field.
The excitation energy of the band-head $1^-$ 
state of the $K^\pi=1^-$ band is predicted to be less than 3 MeV. 
It is surprising that such many-particle many-hole states may exist in 
low-energy region of $^{30}$Ne.
These results indicate the softness of neutron $N=20$ shell,
and point the importance of neutron excitation in the deformed
mean-field in neutron-rich nuclei. 

\begin{figure*}
\epsfxsize=7cm
\centerline{\epsffile{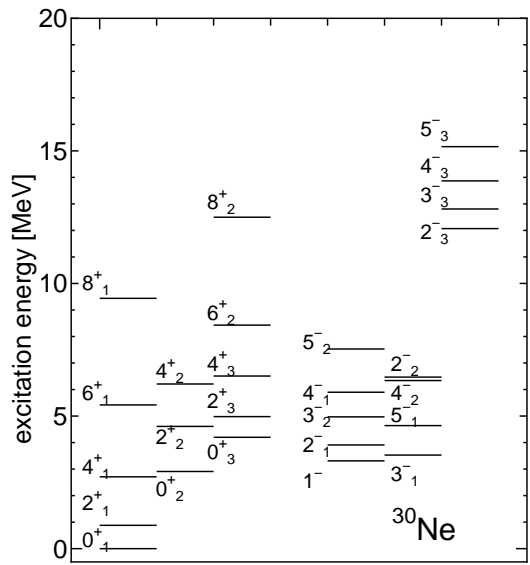}}
\caption{Theoretical energy levels of $^{30}$Ne calculated with 
the AMD+GCM method by Kimura et al. The details of the calculations
are described in \protect\cite{Kimura-ne}.
}
\label{fig:ne}       
\end{figure*}

\section{Three-center clustering and cluster crystallization}
\label{sec:2}

As mentioned before, the $3\alpha$ cluster structure is known in $^{12}$C.
The $3\alpha$ cluster states with a triangular shape 
and linear-chain structure have been
discussed for a long time. Recently, Tohsaki et al. proposed a gas-like dilute
$3\alpha$ state and succeeded to describe the properties of the 
$0^+_2$ state of $^{12}$C with 
Bose-condensed wave functions of 3 $\alpha$ particles. 
In the experimental side, the broad 
resonances at 10 MeV are recently assigned to be $0^+_3$ and $2^+_2$ states,
which are candidates of $3\alpha$-cluster states.

In neutron-rich C isotopes, three-center cluster structures with the 
$3\alpha$-cluster core are expected to exist in the excited states. 
Possible linear chain structures were suggested to appear 
in highly excited states of $^{16}$C and $^{15}$C
\cite{Oertzen97,Itagaki01,Oertzen-rev}.
In $^{14}$C, Itagaki et al. proposed an equilateral-triangular shape
with $3\alpha$ core in $K^\pi=3^-$ band which is stabilized by 
excess neutrons
\cite{Itagaki04}. 
They performed molecular orbital model calculations of $^{14}$C with  
$3\alpha$ and excess neutrons. In their calculations, it was found that
the excess neutrons distribute in the gap space between $\alpha$ cores.
As a result, 
the triangular configuration of $3\alpha$ is stabilized, 
and the $K^\pi=3^-$ rotational band is formed due to the $D_{3h}$
symmetry. The experimentally observed $3^-_2$,
$4^-_1$ and $5^-_1$ states are the candidates of the members of 
this $K^\pi=3^-$ band. 
Comparing the gas-like $3\alpha$ structure in the $0^+_2$ of $^{12}$C, 
the triangular shape of the $3\alpha$ becomes more rigid due to the valence
neutrons in $^{14}$C. Itagaki et al. named this phenomenon as '$\alpha$ 
crystallization'.

\begin{figure}
\epsfxsize=6cm \centerline{\epsffile{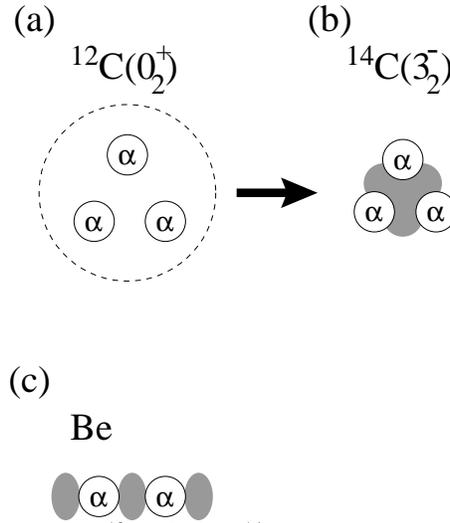}}
\caption{Schematic figures for cluster structure in $^{12}$C$(0^+_2$) (a),
$^{14}$C$(3^-_2$) (b), and neutron-rich Be with the molecular 
$\sigma$-orbitals (c). }
\label{fig:2}       
\end{figure}

The mechanism of rigid cluster structure is understood as follows.
In case of $^{12}$C, 
3 $\alpha$ clusters are weakly bounded and can move freely 
in the cluster states above the $3\alpha$ threshold energy
(Fig.\ref{fig:2}(a)).
When two neutrons are added into those $3\alpha$ states,
the valence neutrons move around the $\alpha$ cores and occupy the 
gap space between $\alpha$ cores(see Fig.\ref{fig:2}(b)). 
As a result, the $\alpha$ clusters can not freely move because the motion
of the $\alpha$ clusters are forbidden due to the Pauli blocking
between the valence neutrons and the neutrons inside the $\alpha$ clusters.
Thus, the 3 $\alpha$ clusters are crystallized to form the triangular shape 
in $^{14}$C.  Itagaki's idea of 
'cluster crystallization' can be also applied to
two-body cluster states as well as the $3\alpha$ cluster states. 
As mentioned before, the developed cluster structures are suggested
in neutron-rich Be and Ne isotopes which have $\alpha+\alpha$-cluster 
and $^{16}$O+$\alpha$-cluster cores, respectively. Especially, 
the remarkable enhancement 
of cluster structure is expected when the valence neutrons occupy the
longitudinal molecular orbitals, namely, the $\sigma$ orbitals.
The $\sigma$ orbitals have nodes along the longitudinal axis.
It is important that the valence neutrons in $\sigma$ orbitals
occupy the gap space between the core clusters(Fig.\ref{fig:2}(c)).
As a result, when the core clusters approach to each other, they feel
repulsion against the valence neutrons in the gap region 
because of Pauli blocking. In other words, due to the existence of the valence
neutron in the gap space, the core clusters can not move so freely and 
are kept away. Thus, the cluster structure is enhanced.

It is concluded that the valence neutrons in molecular orbitals 
play important roles in the cluster states of neutron-rich nuclei.
The valence neutrons bound the clusters more deeply, and may make 
the spatial configuration of clusters more rigid.

\section{Summary}\label{sec:3}
The recent development of theoretical and experimental studies 
revealed that the cluster aspect is an essential feature in unstable
nuclei as well as stable nuclei. The coexistence of cluster and mean-field
aspects brings a variety of structure to unstable nuclei.
We reported some topics concerning the cluster structure of unstable nuclei 
while focusing on the negative parity bands of even-even nuclei.
In the low-lying states of neutron-rich nuclei, the mean-field aspect of 
the valence neutron behavior is found to be essential.
The negative parity rotational bands
appear in the low-energy region due to 
the particle-hole excitation in the deformed neutron mean-field. On the other
hand, in high energy region, there may exist negative parity states 
in the parity doublet bands which are caused by asymmetric intrinsic shapes.
In Be and Ne isotopes, developed cluster structures with 
$2\alpha$-cluster and $^{16}$O+$\alpha$-cluster cores were suggested,
while $3\alpha$ cluster structures were predicted in C isotopes
in many theoretical studies.
The valence neutrons in the molecular orbitals play an important role to
stabilize cluster structure.

%
%

\end{document}
